\documentclass[preprint,showpacs,preprintnumbers,amsmath,amssymb,nofootinbib]{revtex4}

\usepackage{slashed,bbm,nicefrac}

\newcommand{\adj}{\mathop{\mathrm{Adj}}\nolimits}
\newcommand{\tr}{\mathop{\mathrm{Tr}}\nolimits}

\begin{document}
       
\preprint{DESY~11--210\hspace{12.9cm} ISSN 0418--9833}
\preprint{November 2011\hspace{15.5cm}}

\boldmath
\title{Renormalization in general theories with inter-generation mixing}
\unboldmath

\author{Bernd A. Kniehl}
\email{kniehl@desy.de}
\affiliation{II. Institut f\"ur Theoretische Physik,
Universit\"at Hamburg, Luruper Chaussee 149, 22761 Hamburg, Germany}
\author{Alberto Sirlin}
\email{alberto.sirlin@nyu.edu}
\affiliation{Department of Physics, New York University,
4 Washington Place, New York, New York 10003, USA}

\date{\today}

\begin{abstract}
We derive general and explicit expressions for the unrenormalized and
renormalized dressed propagators of fermions in parity-nonconserving theories
with inter-generation mixing.
The mass eigenvalues, the corresponding mass counterterms, and the effect of
inter-generation mixing on their determination are discussed.
Invoking the Aoki-Hioki-Kawabe-Konuma-Muta renormalization conditions and
employing a number of very useful relations from Matrix Algebra, we show
explicitly that the renormalized dressed propagators satisfy important physical
properties.

\end{abstract}

\pacs{11.10.Gh, 11.15.Bt, 12.15.Ff, 12.15.Lk}
\maketitle

\section{Introduction} 

The aim of this paper is to derive general and explicit expressions for the
unrenormalized and renormalized dressed propagators of fermions in
parity-nonconserving theories with inter-generation mixing, and to discuss
their important physical properties and implications.

The results presented here immediately apply to the Standard Theory of 
Elementary Particle Physics, usually referred to as the Standard Model (SM),
as well as its extensions.
As has been known for a long time, the quark fields are subject to
inter-generation mixing, as implemented by the Cabibbo-Kobayashi-Maskawa (CKM)
\cite{Cabibbo:1963yz} quark mixing matrix.
Since neutrino oscillations have been observed experimentally and lower mass
bounds have been established, the lepton fields are known to also undergo
inter-generation mixing.
An early treatment of flavor-changing self-energies, both for leptons and
quarks in bound states, which, however, focuses on finite renormalization
effects, may be found in Ref.~\cite{Donoghue:1979jq}.
On the other hand, our treatment is quite general and takes into account the
full mixing amplitudes.
The renormalization of the CKM matrix has been recently discussed by several
authors;
see, for example, Ref.~\cite{Diener:2001qt} and references cited therein.
Mixing renormalization has also been worked out for theories involving
Majorana neutrinos \cite{Kniehl:1996bd}.

This paper is organized as follows.
Section~\ref{sec:two} discusses the derivation of the unrenormalized dressed
propagators.
The mass eigenvalues, the corresponding mass counterterms, and the effect of
inter-generation mixing on their determination are also analyzed.
Section~\ref{sec:three} discusses the renormalization of the dressed
propagators.
Invoking the Aoki-Hioki-Kawabe-Konuma-Muta (AHKKM) renormalization conditions
and employing very useful relations from Matrix Algebra, it is shown
explicitly that the renormalized dressed propagators satisfy important physical
properties.
Section~\ref{sec:four} contains our conclusions.
The Appendix explains how to derive the two-loop expression for the mass
eigenvalues presented in Sec.~\ref{sec:two}, and how to express the mass
counterterms in terms of the unrenormalized self-energies.

\section{Unrenormalized dressed propagator of mixed fermion system}
\label{sec:two}

As is well known, the unrenormalized mass matrix can be brought to diagonal
form with non-negative eigenvalues by means of bi-unitary transformations on
the left- and right-handed fields.
On this basis, the unrenormalized inverse propagator is\ \ 
$-i I_{ij}(\slashed{p})$, where
\begin{equation}
I_{ij}(\slashed{p})=(\slashed{p}-m_i^0)\delta_{ij}-\Sigma_{ij}(\slashed{p}),
\label{eq1}
\end{equation}
$i$, $j$ are flavor indices\footnote{%
In this paper, repeated indices are not summed, unless a summation symbol is
explicitly included.}
and the self-energies $\Sigma_{ij}(\slashed{p})$ are given by
\begin{equation}
\Sigma_{ij}(\slashed{p})=\left[\slashed{p} (B_{+})_{ij}+(A_{+})_{ij}\right]a_+
+\left[\slashed{p}(B_{-})_{ij}+(A_{-})_{ij}\right]a_- .
\label{eq2}
\end{equation}
In Eq.~(\ref{eq2}), $(A_{\pm})_{ij}$, $(B_{\pm})_{ij}$ are Lorentz-invariant
functions of $p^2$ and $a_{\pm}=(1\pm\gamma_5)/2$ are the chiral
projectors.\footnote{%
Throughout this paper, we adopt the notational conventions of Bjorken and Drell
\cite{bd}.}

Equations (\ref{eq1}) and (\ref{eq2}) can be written in compact form, as
\begin{equation}
 I(\slashed{p})=(\slashed{p} S_+-T_+)a_++(\slashed{p}S_--T_-)a_-,
\label{eq3}
\end{equation}
where $S_+$ and $T_+$ are matrices defined by
\begin{equation}
(S_{\pm})_{ij}=\delta_{ij}-(B_{\pm})_{ij},\qquad
(T_{\pm})_{ij}=m^0_i\delta_{ij}+(A_{\pm})_{ij}.
\label{eq4}
\end{equation}
The unrenormalized dressed propagator is\ \
$iP(\slashed{p})=i(I(\slashed{p}))^{-1}$.\footnote{%
Here and in the following, the matrix $iP(\slashed{p})$ is referred to as the
unrenormalized propagator.
The particle propagators are the elements of this matrix, namely
$iP_{ij}(\slashed{p})$.
An analogous denomination is used in Sec.~\ref{sec:three} for the renormalized
propagator $i\hat{P}(\slashed{p})$ and the renormalized particle propagators
$i\hat{P}_{ij}(\slashed{p})$.}
Writing
$(I(\slashed{p}))^{-1}=(\slashed{p}U_+-V_+)a_++(\slashed{p}U_--V_-)a_-$, we
find the relations
\begin{eqnarray}
S_+V_++T_-U_+&=&0,\label{I}\\
S_-V_-+T_+U_-&=&0,\label{II}\\
p^2S_+U_-+T_-V_-&=&\mathbbm{1},\label{III}\\
p^2S_-U_++T_+V_+&=&\mathbbm{1},\label{IV}
\end{eqnarray}
where $\mathbbm{1}$ stands for the unit matrix.

In order to express $U_{\pm}$ and $V_{\pm}$ in terms of $S_{\pm}$ and
$T_{\pm}$, we first solve for $V_-$ in Eq.~(\ref{II}) and insert the result in
Eq.~(\ref{III}).
This leads to 
\begin{eqnarray}
U_-&=&\left[p^2S_+-T_-(S_-)^{-1}T_+\right]^{-1},\label{V}\\
V_-&=&-(S_-)^{-1}T_+U_-.\label{VI}
\end{eqnarray}
Next we solve for $V_+$ in Eq.~(\ref{I}) and insert the result in
Eq.~(\ref{IV}), which leads to
\begin{eqnarray}
U_+&=&\left[p^2S_--T_+(S_+)^{-1}T_-\right]^{-1},\label{VII}\\
V_+&=&-(S_+)^{-1}T_-U_+.\label{VIII}
\end{eqnarray}
More convenient forms for $U_{\pm}$ are obtained by writing
\begin{eqnarray}
U_-&=&\left[\left(p^2-T_-(S_-)^{-1}T_+(S_+)^{-1}\right)S_+\right]^{-1}
=(S_+)^{-1}(p^2-DC)^{-1},\label{IX}\\
U_+&=&\left[\left(p^2-T_+(S_+)^{-1}T_-(S_-)^{-1}\right)S_-\right]^{-1}
=(S_-)^{-1}(p^2-CD)^{-1},\label{X}
\end{eqnarray}
where
\begin{equation}
C=T_+(S_+)^{-1},\qquad D=T_-(S_-)^{-1}.
\label{eq5}
\end{equation}
It is also convenient to introduce the matrices
\begin{equation}
E=(S_+)^{-1}T_-,\qquad F=(S_-)^{-1}T_+.
\label{eq6}
\end{equation}
Using Eqs~(\ref{V})--(\ref{X}), (\ref{eq5}), and (\ref{eq6}), the
unrenormalized dressed propagator is given by $iP$, where
\begin{equation}
P=(\slashed{p}+E)(S_-)^{-1}(p^2-CD)^{-1}a_++(\slashed{p}+F)(S_+)^{-1}
(p^2-DC)^{-1}a_-,
\label{eq7}
\end{equation}
which is fully expressed in terms of the self-energy matrices $S_{\pm}$ and
$T_{\pm}$.
The matrices $(p^2-CD)^{-1}$ and $(p²-DC)^{-1}$ are related by similarity
transformations, as
\begin{equation}
(p^2-CD)^{-1}=C(p^2-DC)^{-1}C^{-1}=D^{-1}(p^2-DC)^{-1}D.
\label{eq8}
\end{equation}
Writing
\begin{equation}
(p^2-CD)^{-1}=\frac{\alpha_+}{\det(p^2-CD)},\qquad
(p^2-DC)^{-1}=\frac{\alpha_-}{\det(p^2-DC)},
\label{eq9}
\end{equation}
where $\alpha_+$ and $\alpha_-$ are the corresponding adjoint
matrices,\footnote{%
Given a square matrix $M$, in this paper the adjoint matrix $\adj M$ means the
transpose of the matrix whose elements are the cofactors of $M$ (see, for
example, Ref.~\cite{book}.)
We recall that the cofactor $C_{ij}$ of the element $m_{ij}$ of $M$ is
$(-1)^{i+j}$ times the determinant of the matrix obtained by deleting the
$i$-th row and the $j$-th column of $M$.\label{footnote4}}
we see that the determinants are equal and that $\alpha_+$ and $\alpha_-$ are
related by the same similarity transformations as in Eq.~(\ref{eq8}).

Thus, the squared mass eigenvalues $M_i^2$ are the zeros of
$\det(p^2-CD)$, namely they satisfy 
\begin{eqnarray}
\det(M_i^2-Y(M_i^2))&=&0,\label{eq10}\\
Y(p^2)&=&(CD)(p^2)\label{eq11}.
\end{eqnarray}
The off-diagonal elements of $Y(p^2)$ arise from inter-generation mixing and
are, therefore, of ${\cal O}(g^2)$ or higher, where $g$ is a generic
weak-interaction gauge coupling.
As a consequence, if terms of ${\cal O}(g^4)$ are neglected, only the diagonal
elements of $p^2-Y(p^2)$ contribute to the determinant, and the eigenvalues are
of the form
\begin{equation}
\tilde{M}_i^2=\tilde{Y}_{ii}(\tilde{M}_i^2) + {\cal O}(g^4),
\label{eq12}
\end{equation}
where $\tilde{Y}_{ii}(p^2)$ denotes $Y_{ii}(p^2)$ in the absence of
${\cal O}(g^4)$ contributions.
If, instead, terms of ${\cal O}(g^4)$ are retained, but three-loop
contributions and higher are neglected, there are two additional effects:
(i) there are now terms of ${\cal O}(g^4)$ in $Y_{ii}(p^2)$ and
(ii) the non-diagonal elements $Y_{ij}(p^2)$ $(i\neq j)$ contribute to the
determinant.
As a consequence, the mass eigenvalues are now of the form
\begin{equation}
M_i^2=Y_{ii}(M_i^2)+\sum_{j\neq i}\frac{(Y_{ij} Y_{ji})(M_i^2)}{M_i^2-M_j^2}
+{\cal O}(g^6, g^4\alpha_s).
\label{eq13}
\end{equation}
In the Appendix, we outline the derivation of Eq.~(\ref{eq13}) and show how
Eqs.~(\ref{eq4}), (\ref{eq5}), (\ref{eq11}), and (\ref{eq13}) can be used to
express the mass counterterms in terms of the unrenormalized self-energy
functions $A^{\pm}$ and $B^{\pm}$, in the approximation of neglecting
three-loop-contributions.

\section{Renormalized dressed propagator of mixed fermion system}
\label{sec:three}

In order to renormalize $P$ [cf.\ Eq.~(\ref{eq7})], we recall that the
unrenormalized propagator is the Fourier transform of
$\left\langle 0\right| T(\Psi^0(x)\overline{\Psi}^0(0))\left|0\right\rangle$,
where the zero superscripts denote the unrenormalized fields.
In our case, they are column and row fields with components labeled by flavor
indices.
In the following discussion, we assume for simplicity that the fermions are
stable.
Decomposing the fields into right- and left-handed components, as
\begin{equation}
\Psi^0=\Psi_+^0+\Psi_-^0,\qquad
\overline{\Psi}^0=\overline{\Psi}_+^0+\overline{\Psi}_-^0,
\label{eq14}
\end{equation}
where $\Psi_\pm^0=a_\pm\Psi$ and $\overline{\Psi}_\pm^0=\overline{\Psi}a_\mp$,
and taking into account the effect of the chiral projectors $a_{\pm}$, it is
easy to see that the first, second, third, and fourth terms of $P$ arise from
$\left\langle 0\right| T(\Psi_-^0\overline{\Psi}_-^0)\left|0\right\rangle$,
$\left\langle 0\right| T(\Psi_+^0\overline{\Psi}^0_-)\left|0\right\rangle$,
$\left\langle 0\right| T(\Psi^0_+\overline{\Psi}_+^0)\left|0\right\rangle$, and
$\left\langle 0\right| T(\Psi_-^0\overline{\Psi}_+^0)\left|0\right\rangle$,
respectively.

Shifting the fields according to
\begin{eqnarray}
\Psi_+^0&=&Z_+^{\nicefrac{1}{2}}\Psi_+,\qquad
\overline{\Psi}_+^0=\overline{\Psi}_+(Z_+^{\nicefrac{1}{2}})^{\dagger},
\label{eq14a}\\
\Psi_-^0&=&Z_-^{\nicefrac{1}{2}}\Psi_-,\qquad
\overline{\Psi}_-^0=\overline{\Psi}_-(Z_-^{\nicefrac{1}{2}})^{\dagger},
\label{eq14b}
\end{eqnarray}
where $\Psi_{\pm}$ are the renormalized fields, we see that the four terms in
$P$ are multiplied on the left and right by various combinations of
$Z_{\pm}^{\nicefrac{1}{2}}$ and $(Z_{\pm}^{\nicefrac{1}{2}})^{\dagger}$
factors.
Since the time-ordered products are now expressed in terms of renormalized
fields, in order to obtain the renormalized propagator $i\hat{P}$, we must
divide out such factors.
Specifically, the first term in $P$ must be multiplied on the left by
$Z_-^{\nicefrac{-1}{2}}$ and on the right by
$(Z_{-}^{\nicefrac{-1}{2}})^{\dagger}$,
the second term by $Z_{+}^{\nicefrac{-1}{2}}$ on the left and
$(Z_{-}^{\nicefrac{-1}{2}})^{\dagger}$ on the right,
the third term by $Z_{+}^{\nicefrac{-1}{2}}$ on the left and
$(Z_{+}^{\nicefrac{-1}{2}})^{\dagger}$ on the right,
and the fourth term by $Z_{-}^{\nicefrac{-1}{2}}$ on the left and
$(Z_{+}^{\nicefrac{-1}{2}})^{\dagger}$ on the right.

Thus, the renormalized propagator is $i\hat{P}$, where
\begin{eqnarray}
\hat{P}&=&(\slashed{p}Z_-^{\nicefrac{-1}{2}}
+Z_+^{-\nicefrac{1}{2}}E)(S_-)^{-1}(p^2-CD)^{-1}
(Z_-^{\nicefrac{-1}{2}})^{\dagger}a_+
\nonumber\\
&&{}+(\slashed{p}Z_+^{\nicefrac{-1}{2}}
+Z_-^{-\nicefrac{1}{2}}F)(S_+)^{-1}(p^2-DC)^{-1}
(Z_+^{\nicefrac{-1}{2}})^{\dagger}a_-.
\label{eq15}
\end{eqnarray}
Recalling Eqs.~(\ref{eq5}) and (\ref{eq6}), we see that the third and fourth
terms are related to the first and second ones, respectively, by the exchange
$+\leftrightarrow -$.

We now note that the $Z^{\nicefrac{-1}{2}}$ factors in Eq.~(\ref{eq15}) can be
absorbed in a redefinition of the self-energy matrices $S_{\pm}$ and $T_{\pm}$,
namely
\begin{equation}
\hat{S}_{\pm}=(Z_{\pm}^{\nicefrac{1}{2}})^{\dagger}S_{\pm}
Z_{\pm}^{\nicefrac{1}{2}},\qquad
\hat{T}_{\pm}=(Z_{\mp}^{\nicefrac{1}{2}})^{\dagger}T_{\pm}
Z_{\pm}^{\nicefrac{1}{2}}.
\label{eq16}
\end{equation}
Using Eq.~(\ref{eq16}), $\hat{P}$ can be written in the compact form
\begin{equation}
\hat{P}=(\slashed{p}+\hat{E})(\hat{S}_-)^{-1}(p^2-\hat{C}\hat{D})^{-1}a_+
+(\slashed{p}+\hat{F})(\hat{S}_+)^{-1}(p^2-\hat{D}\hat{C})^{-1}a_-
\label{eq17},
\end{equation}
where
\begin{equation}
\hat{C}=\hat{T}_+\hat{S}^{-1}_+,\qquad
\hat{D}=\hat{T}_-\hat{S}^{-1}_-,\qquad
\hat{E}=\hat{S}_+^{-1}\hat{T}_-,\qquad
\hat{F}=\hat{S}_-^{-1}\hat{T}_+.
\label{eq18}
\end{equation}
In particular, $\hat{C}\hat{D}$ and $CD$ are related by a similarity
transformation, as
\begin{equation}
\hat{C}\hat{D}=(Z_-^{\nicefrac{1}{2}})^{\dagger}CD
(Z_-^{\nicefrac{1}{2}})^{\dagger -1},
\end{equation}
so that $\det(p^2-\hat{C}\hat{D})=\det(p^2-CD)$ and the mass eigenvalues are
the zeros of either determinant.
The matrices $\hat{S}_{\pm}$, $\hat{T}_{\pm}$, $\hat{C}$, $\hat{D}$, $\hat{E}$,
and $\hat{F}$ are the renormalized counterparts of $S_{\pm}$, $T_{\pm}$, $C$,
$D$, $E$, and $F$, respectively.

In analogy with Eq.~(\ref{eq8}), we have the relations
\begin{equation}
(p^2-\hat{C}\hat{D})^{-1}=\hat{C}(p^2-\hat{D}\hat{C})^{-1}\hat{C}^{-1}
=\hat{D}^{-1}(p^2-\hat{D}\hat{C})^{-1}\hat{D}.
\label{eq20}
\end{equation}
We note that $\hat{C}\hat{D}$ and $\hat{F}\hat{E}$ are also related by a
similarity transformation and so are $(p^2-\hat{C}\hat{D})^{-1}$ and
$(p^2-\hat{F}\hat{E})^{-1}$, namely
\begin{equation}
\hat{S}_-^{-1}\hat{C}\hat{D}\hat{S}_-=\hat{F}\hat{E},\qquad
\hat{S}_-^{-1}(p^2-\hat{C}\hat{D})^{-1}\hat{S}_-=(p^2-\hat{F}\hat{E})^{-1}.
\label{eq21}
\end{equation}
Interchanging $+\leftrightarrow -$, we obtain
\begin{equation}
\hat{S}_+^{-1}\hat{D}\hat{C}\hat{S}_+=\hat{E}\hat{F},\qquad
\hat{S}_+^{-1}(p^2-\hat{D}\hat{C})^{-1}\hat{S}_+=(p^2-\hat{E}\hat{F})^{-1}.
\label{eq22}
\end{equation}
Using Eqs.~(\ref{eq18}), (\ref{eq20}), (\ref{eq21}), and (\ref{eq22}),
Eq.~(\ref{eq17}) can be cast in the alternative form
\begin{equation}
\hat{P}=a_-(p^2-\hat{F}\hat{E})^{-1}\hat{S}^{-1}_-(\slashed{p}+\hat{C})
+a_+(p^2-\hat{E}\hat{F})^{-1}\hat{S}^{-1}_+(\slashed{p}+\hat{D}).
\label{eq23}
\end{equation}
It differs from Eq.~(\ref{eq17}) in that the chiral projectors $a_{\pm}$ are on
the left side of the expression.
In both Eqs.~(\ref{eq17}) and (\ref{eq23}), the cofactors of $a_-$ and $a_+$
are related by the exchange $+\leftrightarrow -$. 

Writing
\begin{eqnarray}
(p^2-\hat{C}\hat{D})^{-1}&=&\frac{\hat{\alpha}_+}{\det(p^2-\hat{C}\hat{D})},
\qquad
(p^2-\hat{D}\hat{C})^{-1}=\frac{\hat{\alpha}_-}{\det(p^2-\hat{D}\hat{C})},
\label{eq24}\\
(p^2-\hat{F}\hat{E})^{-1}&=&\frac{\hat{\beta}_+}{\det(p^2-\hat{F}\hat{E})},
\qquad
(p^2-\hat{E}\hat{F})^{-1}=\frac{\hat{\beta}_-}{\det(p^2-\hat{E}\hat{F})},
\label{eq25}
\end{eqnarray}
where $\hat{\alpha}_{\pm}$ and $\hat{\beta}_{\pm}$ are the corresponding
adjoint matrices (cf.\ Footnote~\ref{footnote4}), the similarity relations in
Eqs.~(\ref{eq20})--(\ref{eq22}) tell us that
\begin{equation}
\det(p^2-\hat{C}\hat{D})=\det(p^2-\hat{D}\hat{C})
=\det(p^2-\hat{F}\hat{E})=\det(p^2-\hat{E}\hat{F}),
\label{eq26}
\end{equation}
and
\begin{equation}
\hat{S}_{\mp}^{-1}\hat{\alpha}_{\pm}=\hat{\beta}_{\pm}\hat{S}^{-1}_{\mp},
\label{eq27}
\end{equation}
a relation that plays an important r\^ole in our discussion of the propagator's
properties.
We recall that in the previous equations, $\hat{S}_{\pm}$, $\hat{T}_{\pm}$,
$\hat{C}$, $\hat{D}$, $\hat{E}$, $\hat{F}$, $\hat{\alpha}_{\pm}$, and
$\hat{\beta}_{\pm}$ are functions of $p^2$.

We now turn our attention to the renormalization conditions.
As emphasized in the seminal work of AHKKM \cite{Aoki:1982ed}, a fundamental
physical property of the renormalized propagator $i\hat{P}$ is that, as
$\slashed{p}\rightarrow m_n$, where $m_n$ is one of the mass eigenvalues, the
pole $(\slashed{p}-m_n)^{-1}$ should be present only in the diagonal element
$i\hat{P}_{nn}$ of the renormalized propagator matrix.
In order to implement this property, as well as the conventional requirement
that the pole residue equals the imaginary unit, AHKKM proposed suitable
conditions on the renormalized inverse propagators, which were described both
graphically and mathematically.

Recalling Eq.~(\ref{eq3}), in our general matrix notation the renormalized
inverse propagator is $-i\hat{I}(\slashed{p})$, where
\begin{equation}
\hat{I}(\slashed{p})=(\slashed{p}\hat{S}_+-\hat{T}_+)a_+
+(\slashed{p}\hat {S}_--\hat{T}_-)a_- .
\label{eq28}
\end{equation}
An alternative expression is
\begin{equation}
\hat{I}(\slashed{p})=a_-(\slashed{p}\hat{S}_+-\hat{T}_-)
+a_+(\slashed{p}\hat{S}_--\hat{T}_+),
\label{eq29}
\end{equation}
where the chiral projectors $a_{\pm}$ are placed on the left.
The homogeneous AHKKM renormalization conditions read
\begin{eqnarray}
\overline{u}_n(\slashed{p})\hat{I}_{nl}(\slashed{p})&=&0,\label{eq30}\\
\hat{I}_{ln}(\slashed{p})u_n(\slashed{p})&=&0\label{eq31},
\end{eqnarray}
where $u_n(\slashed{p})$ is a spinor that satisfies
$\slashed{p}u_n(\slashed{p})=m_nu_n(\slashed{p})$,
$\overline{u}_n(\slashed{p})$ its hermitian adjoint, and $n$ and $l$ are flavor
indices.

Inserting Eq.~(\ref{eq28}) into Eq.~(\ref{eq30}), we have
\begin{equation}
\left[m_n\hat{S}_{\pm}(m_n^2)-\hat{T}_{\pm}(m_n^2)\right]_{nl}=0.
\label{eq32}
\end{equation}
Multiplying on the right by $\hat{S}_{\pm}^{-1}(m_n^2)_{lj}$, summing over $l$,
and remembering the definitions in Eq.~(\ref{eq18}), this becomes
\begin{equation}
\hat{C}_{nj}(m_n^2)=\hat{D}_{nj}(m_n^2)=m_n\delta_{nj},
\label{eq33}
\end{equation}
which implies
\begin{equation}
(\hat{C}\hat{D})_{nn}(m_n^2)=m_n^2,\qquad
(\hat{C}\hat{D})_{nj}(m_n^2)=0\qquad (j\neq n),
\label{eq34}
\end{equation}
with the analogous result for $(\hat{D}\hat{C})(m_n^2)$.

Inserting Eq.~(\ref{eq29}) into Eq.~({\ref{eq31}), recalling the definitions in
Eq.~(\ref{eq18}), and carrying out the analogous analysis, we obtain
\begin{equation}
\hat{E}_{in}(m_n^2)=\hat{F}_{in}(m_n^2)=m_n\delta_{in},
\label{eq35}
\end{equation}
which leads to
\begin{equation}
(\hat{E}\hat{F})_{nn}(m_n^2)=m_n^2,\qquad
(\hat{E}\hat{F})_{in}(m_n^2)=0\qquad (i\neq n),
\label{eq36}
\end{equation}
and the analogous result for $(\hat{F}\hat{E})(m_n^2)$.

Equation~(\ref{eq34}) tells us that, as $p^2\rightarrow m_n^2$, all the
elements in the $n$-th row of $p^2-\hat{C}\hat{D}$ and $p^2-\hat{D}\hat{C}$
vanish.
Therefore, the only non-vanishing cofactors of $(p^2-\hat{C}\hat{D})$ and
$(p^2-\hat{D}\hat{C})$ are those corresponding to the elements of that row,
namely the cofactors $C_{nl}$.
Since the adjoint matrices are the transpose of the cofactor matrices (cf.\
Footnote~\ref{footnote4}), we conclude that the only non-vanishing elements of
$\hat{\alpha}_{\pm}(m_n^2)$ are those in the $n$-th column, namely the elements
$(\hat{\alpha}_{\pm})_{in}(m_n^2)$.
Similarly, from Eq.~(\ref{eq36}) we see that, as $p^2\rightarrow m_n^2$, all
the elements in the $n$-th column of $p^2-\hat{F}\hat{E}$ and
$p^2-\hat{E}\hat{F}$ vanish.
Consequently, the only non-vanishing elements of $\hat{\beta}_{\pm}(m_n^2)$ are
those in the $n$-th row, namely $(\hat{\beta}_{\pm})_{nj}(m_n^2)$.
In combination with Eq.~(\ref{eq27}), these results imply that, as
$p^2\rightarrow m_n^2$, the only non-vanishing elements of the matrices
$\hat{S}_{\mp}^{-1}\hat{\alpha}_{\pm}$ and
$\hat{\beta}_{\pm}\hat{S}^{-1}_{\mp}$ are the diagonal $nn$ elements
$(\hat{S}_{\mp}^{-1}\hat{\alpha}_{\pm})_{nn}(m_n^2)
=(\hat{\beta}_{\pm}\hat{S}^{-1}_{\mp})_{nn}(m_n^2)$.
Thus,
\begin{equation}
(\hat{S}_{\mp}^{-1}\hat{\alpha}_{\pm})_{ij}(m_n^2)
=(\hat{\beta}_{\pm}\hat{S}_{\mp}^{-1})_{ij}(m_n^2)=0\qquad
(i\text{ or } j\neq n).
\label{eq37}
\end{equation}
To examine the effect of these results on the renormalized propagators, we
insert Eq.~(\ref{eq24}) and Eq.~(\ref{eq25}) into Eq.~(\ref{eq17}) and
Eq.~(\ref{eq23}), respectively.
Recalling Eq.~(\ref{eq26}), we obtain
\begin{equation}
\hat{P}=\frac{(\slashed{p}+\hat{E})(\hat{S_-})^{-1}\hat{\alpha}_+a_+
+(\slashed{p}+\hat{F})(\hat{S}_+)^{-1}\hat{\alpha}_-a_-}{%
\det(p^2-\hat{C}\hat{D})}
\label{eq38}
\end{equation}
from Eq.~(\ref{eq17}), and
\begin{equation}
\hat{P}=\frac{a_-\hat{\beta}_+(\hat{S_-})^{-1}(\slashed{p}+\hat{C})
+a_+\hat{\beta}_-(\hat{S}_+)^{-1}(\slashed{p}+\hat{D})}{%
\det(p^2-\hat{F}\hat{E})}
\label{eq39}
\end{equation}
from Eq.~(\ref{eq23}).

Using Eqs.~(\ref{eq33}), (\ref{eq35}), and (\ref{eq37}), one readily verifies
that, as $p^2\rightarrow m_n^2$, the only non-vanishing elements in the
numerators of Eqs.~(\ref{eq38}) and (\ref{eq39}) are, in fact, the diagonal
$nn$ elements.
Thus, the explicit expressions of the renormalized propagator $i\hat{P}$, given
in Eqs.~(\ref{eq17}), (\ref{eq23}), (\ref{eq38}), and (\ref{eq39}), indeed
satisfy the fundamental physical property that the $(\slashed{p}-m_n)^{-1}$
pole is present only in the diagonal element $i\hat{P}_{nn}$ of the propagator
matrix.

The inhomogeneous AHKKM renormalization conditions are
\begin{eqnarray}
\frac{1}{\slashed{p}-m_n}\hat{I}_{nn}(\slashed{p})u_n(\slashed{p})
=u_n(\slashed{p}),
\label{eq40}\\
\overline{u}_n(\slashed{p})\hat{I}_{nn}(\slashed{p})\frac{1}{\slashed{p}-m_n}
=\overline{u}_n(\slashed{p}).
\label{eq41}
\end{eqnarray}
Inserting Eq.~(\ref{eq28}) into Eq.~(\ref{eq40}), expanding the numerator about
$\slashed{p}=m_n$, and using Eq.~(\ref{eq32}), we find the renormalization
conditions
\begin{eqnarray}
\left\{\hat{S}_-(m_n^2)+m_n\left[m_n(\hat{S}_++\hat{S}_-)
-\hat{T}_+-\hat{T}_-\right]^\prime\right\}_{nn}&=&1,
\nonumber\\
(\hat{S}_-)_{nn}(m_n^2)&=&(\hat{S}_+)_{nn}(m_n^2),
\label{eq42}
\end{eqnarray}
where the prime symbol stands for the derivative with respect to $p^2$,
evaluated at $p^2=m_n^2$.
Inserting Eq.~(\ref{eq29}) into Eq.~(\ref{eq41}), we obtain the same result.

In order to analyze the effect of Eq.~(\ref{eq42}), we evaluate the residue of
the $(\slashed{p}-m_n)^{-1}$ pole in $\hat{P}$ using Eq.~(\ref{eq38}) and focus
on the $a_+$ term.
We expand $\det(p^2-\hat{C}\hat{D})$ about $p^2=m_n^2$ through
${\cal O}(p^2-m_n^2)$.
Since $p^2=m_n^2$ is a zero of the determinant, the first term vanishes, and we
have
\begin{equation}
\det(p^2-\hat{C}\hat{D})
=\left[\det(p^2-\hat{C}\hat{D})\right]^\prime(p^2-m_n^2)+\cdots.
\label{eq43}
\end{equation}
Using the well known expression
\begin{equation}
(\det M)^\prime=\tr\left(M^\prime\adj M\right),
\label{eq44}
\end{equation}
the r.h.s.\ of Eq.~(\ref{eq43}) becomes
$\tr\left\{\hat{\alpha}_+(m_n^2)\left[1-(\hat{C}\hat{D})^\prime\right]\right\}
(p^2-m_n^2)+\cdots$.
Multiplying by $\slashed{p}-m_n$, taking the limit
$\slashed{p}\rightarrow m_n$, and recalling Eqs.~(\ref{eq35}) and (\ref{eq37}),
we see that the residue of the $(\slashed{p}-m_n)^{-1}$ pole in the $a_+$ term
of Eq.~(\ref{eq38}) is
\begin{equation}
\text{Res}_+=\frac{(\hat{S}^{-1}_-\hat{\alpha_+})_{nn}}{
\tr\left\{\hat{\alpha}_+\left[1-(\hat{C}\hat{D})^\prime\right]\right\}}.
\label{eq45}
\end{equation}
Here and in the following, it is understood that all the functions are
evaluated at $p^2=m_n^2$.
To simplify this expression, we insert $\hat{S}_-\hat{S}_-^{-1}=1$ in the
argument of the trace.
Recalling again Eq.~(\ref{eq37}), we find
\begin{equation}
\tr\left\{\hat{S}_-\hat{S}_-^{-1}\hat{\alpha}_+
\left[1-(\hat{C}\hat{D})^\prime\right]\right\}
=\left(\hat{S}_-^{-1}\hat{\alpha}_+\right)_{nn}
\left\{\left[1-(\hat{C}\hat{D})^\prime\right]\hat{S}_-\right\}_{nn},
\end{equation}
and the residue becomes
\begin{equation}
\text{Res}_+=\frac{1}{\left\{\left[1-(\hat{C}\hat{D})^\prime\right]
\hat{S}_-\right\}_{nn}}.
\label{eq47}
\end{equation}

Taking into account Eqs.~(\ref{eq33})--(\ref{eq35}), Eq.~(\ref{eq47}) becomes
\begin{equation}
\text{Res}_+=\frac{1}{\left\{\hat{S}_-+m_n\left[m_n(\hat{S}_++\hat{S}_-)
-\hat{T}_+-\hat{T}_-\right]^\prime\right\}_{nn}}.
\label{eq48}
\end{equation}
Thus, the renormalization condition of Eq.~(\ref{eq42}) indeed implies that
\begin{equation}
\text{Res}_+=1.
\label{eq49}
\end{equation}
Calling $\text{Res}_-$ the residue of the $(\slashed{p}-m_n)^{-1}$ pole in the
$a_-$ term of Eq.~(\ref{eq38}), an analogous analysis shows that
\begin{equation}
\text{Res}_-=1.
\label{eq50}
\end{equation}
We conclude that, when the inhomogeneous renormalization condition of
Eq.~(\ref{eq40}) is imposed, the poles in our explicit expressions for the
renormalized propagator [cf.\ Eqs.~(\ref{eq17}), (\ref{eq23}), (\ref{eq38}),
and (\ref{eq39})] have residues $i$.

\section{Conclusions} 
\label{sec:four} 

We derived general and explicit expressions for the unrenormalized and
renormalized dressed propagators of fermions in parity-nonconserving theories
with inter-generation mixing [cf.\ Eqs.~(\ref{eq7}), (\ref{eq17}),
(\ref{eq23}), (\ref{eq38}), and (\ref{eq39})].
We analyzed the determination of the mass eigenvalues and the corresponding
mass counterterms in the approximation of neglecting three-loop contributions
[cf.\ Eqs.~(\ref{eq13}) and (\ref{eqa9})].
In particular, we discussed the effect of inter-generation mixing on these
determinations.
Using the AHKKM renormalization conditions and applying very useful relations
from Matrix Algebra, we showed explicitly that our renormalized dressed
propagator [cf.\ Eqs.~(\ref{eq17}), (\ref{eq23}), (\ref{eq38}), and
(\ref{eq39})], which is valid to all orders in perturbation theory, satisfies
important physical properties.
In turn, this demonstrates in a clear manner that the AHKKM renormalization
conditions are also valid to any order of perturbation theory.

\begin{acknowledgments}
This work was supported in part by the German Research Foundation through the 
Collaborative Research Center No.~676 {\it Particles, Strings and the Early
Universe --- The Structure of Matter and Space Time}.  
The work of A. Sirlin was supported in part by the National Science Foundation 
through Grant No.\ PHY--0758032. 
\end{acknowledgments}

\begin{appendix}

\section{Appendix}

In this appendix, we outline the derivation of Eq.~(\ref{eq13}) in the
approximation of neglecting three-loop contributions and show how it can be
applied to express the mass counterterms in terms of the basic self-energy
functions $(A_\pm)_{ij}$ and $(B_\pm)_{ij}$ in Eq.~(\ref{eq2}).
For simplicity, we consider the three-generation case.

As explained in the paragraph containing Eqs.~(\ref{eq10}) and (\ref{eq11}),
the mass eigenvalues are the zeros of $\det(p^2-Y(p^2))$, where
$Y(p^2)=(CD)(p^2)$ and the matrices $C$ and $D$ are defined in Eq.~(\ref{eq5}).
Using Eqs.~(\ref{eq4}), (\ref{eq5}), and (\ref{eq11}), we find
\begin{equation}
Y=(M^0)^2+Z,
\label{eqa1}
\end{equation}
where $M^0$ is the diagonal bare mass matrix with elements $m_i^0$ and
\begin{eqnarray}
Z&=&(M^0)^2B_-(1+B_-)+M^0(A_-+A_-B_-+B_+A_-)+M^0B_+(1+B_+)M^0
\nonumber\\
&&{}+A_+(1+B_+)M^0+A_+M^0B_-+M^0B_+M^0B_-+A_+A_-.
\label{eqa2}
\end{eqnarray}
In Eq.~(\ref{eqa1}), we have separated out the squared bare mass term
$(M^0)^2$ and the one- and two-loop contributions contained in $Z$.
We recall that $A_\pm$, $B_\pm$, $Y$, and, consequently, $Z$ are functions of
$p^2$.
It is further convenient to split
\begin{equation}
(M^0)^2=M^2+\delta M^2,
\label{eqa3}
\end{equation}
where $M$ is the renormalized mass matrix whose elements are the mass
eigenvalues and $\delta M^2$ is the mass counterterm matrix.
Thus,
\begin{equation}
Y=M^2+X,
\label{eqa4}
\end{equation}
where
\begin{equation}
X=\delta M^2+Z.
\end{equation}
We note that, in Eq.~(\ref{eqa4}), $M^2$ contains the zeroth-order terms, while
$X$ contains the one- and two-loop contributions.

Neglecting three-loop contributions, in the three-generation case the
eigenvalue equation $\det(p^2-Y(p^2))=0$ becomes
\begin{equation}
(p^2-Y_{11})(p^2-Y_{22})(p^2-Y_{33})-(p^2-Y_{11})Y_{23}Y_{32}
-(p^2-Y_{22})Y_{13}Y_{31}-(p^2-Y_{33})Y_{12}Y_{21}=0.
\end{equation}
Consider the neighborhood of $p^2=M_1^2$, where $M_1$ is one of the mass
eigenvalues: dividing by $(p^2-Y_{22})(p^2-Y_{33})$, we have
\begin{equation}
(p^2-Y_{11})\left[1-\frac{Y_{23}Y_{32}}{(p^2-Y_{22})(p^2-Y_{33})}\right]
=\frac{Y_{13}Y_{31}}{p^2-Y_{33}}+\frac{Y_{12}Y_{21}}{p^2-Y_{22}}.
\label{eqa7}
\end{equation}
The factors $Y_{23}Y_{32}$, $Y_{13}Y_{31}$, and $Y_{12}Y_{21}$ are of two-loop
order or higher.
As $p^2\to M_1^2$, we see from Eq.~(\ref{eqa4}) that, to leading order, we
have $p^2-Y_{22}=M_1^2-M_2^2$ and $p^2-Y_{33}=M_1^2-M_3^2$.
Thus, neglecting three-loop contributions, as $p^2\to M_1^2$, Eq.~(\ref{eqa7})
reduces to
\begin{equation}
M_1^2=Y_{11}(M_1^2)+\frac{(Y_{12}Y_{21})(M_1^2)}{M_1^2-M_2^2}
+\frac{(Y_{13}Y_{31})(M_1^2)}{M_1^2-M_3^2},
\end{equation}
which is a particular case of Eq.~(\ref{eq13}).

Recalling Eqs.~(\ref{eq13}), (\ref{eqa1}), and (\ref{eqa3}), the mass
counterterms are then
\begin{eqnarray}
\delta M_i^2&=&(m_i^0)^2-M_i^2
\nonumber\\
&=&(m_i^0)^2-Y_{ii}(M_i^2)
-\sum_{j\ne i}\frac{(Y_{ij}Y_{ji})(M_i^2)}{M_i^2-M_j^2}
\nonumber\\
&=&-Z_{ii}(M_i^2)
-\sum_{j\ne i}\frac{(Z_{ij}Z_{ji})(M_i^2)}{M_i^2-M_j^2},
\label{eqa9}
\end{eqnarray}
where $Z$ is defined in Eq.~(\ref{eqa2}).
In the last equality of Eq.~(\ref{eqa9}), we have replaced $Y_{ij}\to Z_{ij}$,
since both are equal when $i\ne j$ [cf.\ Eq.~(\ref{eqa1})].

We note that, subject to our approximation, the amplitudes involving linear
powers of $A_\pm$ and $B_\pm$ in Eq.~(\ref{eqa2}) contain both one- and
two-loop contributions.

Using Eq.~(\ref{eqa2}), we find for the diagonal terms
\begin{eqnarray}
Z_{ii}&=&(m_i^0)^2(B_++B_-+B_+^2+B_-^2)_{ii}
+m_i^0(A_++A_-+A_+B_++A_-B_-+B_+A_-)_{ii}
+(A_+A_-)_{ii}
\nonumber\\
&&{}+\sum_{j=1}^3\left[m_j^0(A_+)_{ij}(B_-)_{ji}+m_i^0m_j^0(B_+)_{ij}(B_-)_{ji}
\right].
\label{eqa10}
\end{eqnarray}
We note that $Z_{ii}$ depends not only on the bare fermion masses $m_i^0$ and
$m_j^0$ displayed in Eq.~(\ref{eqa10}), but also on additional ones present in
the loop diagrams.
We refer generically to the latter as $m_l^0$.
Consistently with our approximation, in the contributions of two-loop order, we
replace the bare masses  $m_i^0$, $m_j^0$, and $m_l^0$ by the mass eigenvalues
$M_i$, $M_j$, and $M_l$, respectively.
In the contributions of one-loop order, we replace
\begin{equation}
m_i^0=\left[M_i^2-Z_{ii}^{(1)}(M_i^2)\right]^{\nicefrac{1}{2}},
\label{eqa11}
\end{equation}
and similarly for $m_l^0$.
In Eq.~(\ref{eqa11}), the superscript $(1)$ stands for the one-loop
contribution, namely
\begin{equation}
Z_{ii}^{(1)}=M_i^2(B_+^{(1)}+B_-^{(1)})(M_i^2)
+M_i(A_+^{(1)}+A_-^{(1)})(M_i^2),
\end{equation}
with an analogous expression for $Z_{ll}^{(1)}$.

The contributions involving $Z_{ij}Z_{ji}$ with $j\ne i$ in Eq.~(\ref{eqa9})
are already of two-loop order or higher, so that in the off-diagonal amplitudes
$Z_{ij}$ with $j\ne i$, we simply replace $m_i^0,m_j^0,m_l^0\to M_i,M_j,M_l$.
In this way, subject to the approximation of neglecting three-loop
contributions, the mass counterterms $\delta M_i^2$ given in Eq.~(\ref{eqa9})
are fully expressed in terms of the basic self-energies $A_\pm(M_i^2)$ and
$B_\pm(M_i^2)$ of Eq.~(\ref{eq2}) and the mass eigenvalues.

\end{appendix}

\end{document}